# Planetary dependence of *melanoma*.

Konstantin Zioutas [1,*] and Edward Valachovic [2]

[1*] Physics Department, University of Patras, Patras, Greece;
*Corresponding author*, *Email*: zioutas@cern.ch .
[2] Department of Epidemiology and Biostatistics. School of Public Health. University at Albany, State University of New York, New York, USA;
*Email*: evalachovic@albany.edu .

**Abstract:**
We present signature for planetary correlations following an analysis of monthly melanoma rates in USA for the period 1973 - 2011. Apart from seasonal observations, to the best of our knowledge, a planetary relationship in medicine is observed for the first time. The statistical significance is well above 5σ, while various crosschecking make systematics highly improbable as the cause. The observed planetary dependence in physics was suggestive for this investigation, which extends the otherwise unexpected planetary connection. Thus, streaming invisible matter from the dark sector, whose flux can be occasionally enhanced towards the Earth via planetary gravitational focusing, and, even much stronger by the Sun, it may be the explanation for 1-10% of melanoma diagnoses. The derived shortest melanoma periodicity of about 87.5 days points *in its own right* at a short latency period of about few months. Contrariwise, the present findings strengthen the previous physics claim of streams of invisible matter.

## 1. Introduction

There are unsolved problems in physics and medicine, which conventional approaches failed to localize an origin as this lasts for very long time. The driving question behind this investigation is as to whether mysteries in physics and medicine have as common origin the dark Universe we are living in. In a recent work [1] it has been argued observationally that streaming "invisible massive matter" gravitationally focused by the Planets and/or the Sun are the cause of persisting anomalies of solar activity and the excess degree of ionization of the Earth's ionosphere. Interestingly, the Sun is by far the best performing gravitational lens than any planet of the solar system (up to about $10^4$ to $10^5$ times). The planetary dependence found in ref. [1] was not expected within known physics. Since the alternative solutions by planetary tidal gravitational interactions or the magnetic fields are far too weak to cause any effect over large distances, and, in particular, to have significant changes within time intervals being much shorter than their orbital period.

The aim of this work is to find out, for the first time, to the best of our knowledge, whether planetary correlations exist also in medical data. For this purpose we started with high statistics data sets on the appearance rate of *melanoma*. Needless to say that cancer is among the biggest mysteries in medicine, while the nature of the Dark Universe is an intriguing open question in Physics. Following the present findings of this investigation along with those of ref. [1]: if both decoupled every day's distinct observations have nevertheless common origin, the additional information from this investigation may help to identify the cause of both persisting mysteries.

## 2. Observation - Analysis

In this work we used monthly Melanoma Rates normalised to a population of 100000 which have been obtained from SEER, the Surveillance Epidemiology End Results database [2]. Figure 1 shows the "light curve" of diagnosed melanoma in the period 1/1/1973 to 31/12/2010 (38 years). The striking annual oscillatory behavior is seen even in the raw data. Triggered by this planetary observation, i.e., the peaking seasonal rate, here we go in depth of this first planetary hint. For this, each monthly mean value divided by the length of the corresponding month has been equally distributed over the days of each month. Similarly, we have used also a linear interpolation between neighboring monthly values. Both approximations cannot cause significant artifacts as it is demonstrated throughout this work (see in particular Figures 1,3,5,6,8,9). However, using large BIN-values in longitude, e.g. 18º or 30º, the original seasonal distribution is almost recovered (e.g., compare the insert in Figure 1 with Figure 7a (37x12)). In other words, using large BIN values the daily splitting approximation is cancelled to a high degree. Note, in this work, if for the daily splitting the linear interpolation is used, it is mentioned explicitly.

The diagnosed number of melanoma was about 10000 / year and 30000 / year at the beginning and the end of the 38 years observation, i.e., the total number of diagnoses is about 760000. For a spectrum with 10 points, the standard deviation for Poisson statistics is equal to 0.36% per point. In order to have >5σ statistical significance between two points, a 2.6% difference is sufficient. It is worth mentioning that for the statistical treatment we use only the raw data, i.e., the original SEER monthly values.

Two independent analysis methods were used: a) by projecting the day of melanoma diagnosis on the corresponding planetary longitude [3], or, by building the sum of specific consecutive time intervals day by day [4]. In both cases, eccentricity related effects are finally factored-out, because we normalize to 1 day.

For the present planetary related search [1], we also focus on the three inner solar planets due to their short orbital periods. This implies many orbits during the 38 years of the observation:

157 for Mercury (88 days), 61 for Venus (225 days) and 38 for Earth. Any unnoticed temporal spike will be diminished accordingly. Or, to put it differently, observing a longitudinal clustering of events after so many repeating periods, it cannot be random in origin. In order to further exclude systematics, we split the integrated spectra of Venus and Mercury in two halves and compare them (see Figures 2-3, but also 9-13). This is an important consistency test. In the case of the Earth, the effect is so big (>30% in amplitude) that it is visible every year even in the raw data (see Figure 1).

## 2a. Planetary longitudinal distributions

The projected longitudinal distribution of the derived daily melanoma rate (using linear interpolation) for the whole time interval of 38 years for the inner 3 planets (Earth, Venus and Mercury) is shown in Figures 1-3 [3]. All figures with their captions are self-explanatory. In order to crosscheck the whole evaluation procedure, e.g., for Venus, the same computer program was running by replacing the SEER data by a constant random event rate of 1/day. The result is, as expected, a completely flat distribution at 1/day (Figure 2 A "SEER"). In addition, the 61 Venus orbits have been split in two equal halves (31 and 30 orbits). Interestingly, the spectral shape and the corresponding amplitude does not show any essential difference in all 3 spectra (Figure 2). However, the little different spectral shapes in Figure 2B could point at an accumulated phase-shift during the 31 orbits between both halves of about +60º (~40 days / 31 orbits). Such a phase shift would point at a real data periodicity of 224.47 - 40/31≈223.2 days (see also Figure 6). This periodicity increases if neighboring synodic periods of Venus (225.2 to 237 days) play a role (see next sub-section and Figure 6). Given the similarity between the two spectra in Figure 2B, no firm conclusion about the actual data periodicity near 224.47 days can be made. However, this goes beyond the purpose of this work, which is to establish a planetary dependence in medical data. Relevant is the observed amplitude of 4.5% of the Venus orbital spectrum.

Similarly, Figure 3 shows the projected longitudinal distribution of melanoma event rate along Mercury's orbit. Here, the 157 Mercury orbits are split in two halves (79 and 78 orbits), showing similar spectral shapes, while the amplitudes are rather equal to each other, excluding also here an unnoticed temporal spike as the cause. The accumulated phase shift during the 79 orbits is about 120º to 150º, or about 40 days in 79 orbits between both halves (Figure 3B). This phase shift points at an underlying periodicity near 88 days, i.e., 88-40/79≈87.5 days (see Figure 6). At present, to precisely explain such a small phase mismatch it goes beyond the scope of this work, which is to establish a planetary dependence. We recall that from the 8 synodic periods of Mercury, 6 have a period between 88.09 and 101 days, which cause a phase shift, but in the opposite direction.

To further support a planetary dependence of the melanoma rate following the SEER data, Figure 4 shows combined longitudinal distributions of the inner 3 planets by requiring for each additional planetary constraints. It is worth noticing that when comparing a pair, what counts are possible differences between: their spectral shape, the amplitude of the accumulated SEER rate (=melanoma rate) as well as the longitudinal position of peaking distributions. In fact, these plots, and in particular the peaking ones, strengthen the results from the single planetary longitudinal distributions of Figure 2 and 3. Because, any long range force extends smoothly during a planetary orbit, which is not the case in some of the spectral shapes in Figure 4.

SIMULATION: In order to validate the evaluation method using the daily splitting, we have used a list of daily flares from 1972 to 2015 (provided by M.J. Aschwanden). They were obtained with an automated flare detection algorithm, which is 5 times more sensitive than the official NOAA flare catalog [5]. Their longitudinal distribution gives even better peaks in all inner 3 planetary distributions compared with the corresponding ones in ref. [1]. Then, the number of flares of 30 consecutive days are summed-up, in order to mimic the monthly rate of the SEER data [2]. Following the aforementioned linear interpolation algorithm, the daily distribution of flares between the simulated neighboring monthly number of flares is estimated. Then, running the same analysis code with the daily split solar flares, the so simulated spectral distributions are derived (Figure 5, *right*). For the purpose of this work, the observed degree of distortion of the original spectral shapes (Figure 5, *left*) is acceptable, justifying thus the approximation with the introduced daily splitting, even though this procedure leads to a worse signal-to-noise ratio (maximum for Mercury by a factor ~2.5×). Note, the BIN is larger than the corresponding one with the real data. Therefore, we are confident to include in the analysis all inner planets. Even for Mercury, with the shortest orbital period of 88 days, the initial bell-shaped distribution survives, indicating also, to some degree, the re-appearance of peaks present in the initial spectrum. In fact, both splitting algorithms result to quite similar results, though short oscillations are filtered out by the linear interpolation algorithm (not shown). We stress that in this work the significance of not random distributions is based exclusively on raw data (see Figures 7).

## 2b. Periodic distributions

For crosschecking reasons, the results in this section are derived using a completely different evaluation concept [4], which allows to search for *any* periodic distribution in the SEER data set. Figure 6 explains the method by choosing, for example, a period of 360 days, while the melanoma rate follows an annual rhythm (Figure 1). For example, the reconstructed daily melanoma rate of consecutive time intervals of 360 days is added, day by day, and this repeats 38 times during the observation time 1973-2011. Figure 6 (*left*) gives the so derived spectrum, showing also how smoothly the distribution becomes after stacking 38 cycles of 360 days, whose period differs only by 1.4% from the actual annual period (Figure 1). Then, the same spectrum is calculated for two half time intervals of 19 years (Figure 6, *right*). An inspection of both halves shows that they are phase-shifted relative to each other. In fact, the observed phase shift (about 105 days) is reproduced by the accumulated phase difference for every year, which is equal to 19×(365.25-360) ≈ 100 days. This phase-shift provides also an interesting information: if it is not zero (ideal case), then the underlying periodicity is not the chosen one (i.e., the 360 days in this example), but that around 365 days as we could recover it from the observed phase-shift and its sign. In other words, by plotting even only two halves, the actual periodicity which causes the phase mismatch can be identified. *This is an important finding of potential interest, as for datasets with short time BIN it allows to identify the underlying periodic impact.*

Interestingly, beyond the mentioned synods for Mercury and Venus, the solar system has some 30 orbital and synodic periodicities plus several multiples thereof (see Figures 7-11). Widening known planetary periods due to the aforementioned interference effect with near periodicities (see Figure 6), the interval up to about 850 days is quite crowded with planetary periodicities. Thus, for this work, unbiased potential periodicities of interest are: the planetary orbital periods (Figure 2,3,7,8), the synodic periods between any planets (Figure 8,10,11) and multiples thereof (Figure 7,8,9). In addition, in searching with this method in a large data set of sunspots [6] some periods of as yet unidentified origin provided striking spectral shape (see e.g. Figure 9, 12). Moreover, high energy solar flares show the "Rieger periodicity" around 156 days [8] of unknown origin (see Figure 13). Thus, following such criteria, we have selected periods which are observationally motivated, covering a wide range of the planetary and the active Sun rhythmic behavior. Following the reasoning of this Section, more results were thus obtained, which are given self-explanatory in Figures 7-13.

RAW DATA ANALYSIS: Figure 7a shows the derived evaluation results by using the original monthly values of the SEER data. This allows us to precisely estimate the statistical significance of a modulation amplitude, which, in the absence of planetary impact, it should be randomly distributed. We begin with the summing-up the 1/3/1973, in order to start at the minimum of the oscillatory behavior, losing thus only 2 months out of 456. It is clearly visible a kind of fundamental period of ~3 months, which is approximately equal to the Mercury's orbital period. It is worth mentioning that the reconstructed spectral shape for a period of 12 months, and, the multiples thereof (24 and 36 months), may be seen as validation of the applied evaluation procedure of the raw data. In addition, these spectra serve as a crosscheck of the applied evaluation procedure using the estimated daily melanoma rate, and *vice versa*. In fact, they have provided a fundamental periodicity at 87.5 days (see subsection 2a. and Figure 3).

In addition, from the 12 spectra shown in Figure 7a, we did single out two with a period of 21 and 27 months (Figure 7b). The comparison with nearby time intervals is instructive. For example, in the case of a period of 21 months, the remarkable 7 times repeating "resonance" disappears in both neighboring time intervals. Along this reasoning, in the case of 27 months summing-up mode with striking 9 peaks, the Fourier analysis recovers the aforementioned 'fundamental' period at ~3 months, which is quite suppressed in the Fourier power spectrum of the nearby interval of 26 months (see inserts in Figure 7b). For the time interval of 26 months, an apparent irregular oscillatory behavior is seen. Assuming Poisson statistics, the relative error per point is <0.7%. The estimated significance for the amplitude of 7% shown in Figure 7b is about **7σ**.

## 2c. Fourier analysis

For a completely independent crosschecking, a Fourier analysis was performed using the original 456 monthly SEER values (1973-2010). Figure 7c gives the periodogram. The two inserts zoom at the regions around 90 days and 225 days, i.e., near the orbital periods of Mercury and Venus, respectively. The 91.3 days peak is the 4$^{th}$ harmonic of the main annual peak, while the origin of the ~3.5σ peak at 220 days is unknown. Noticeably, there is a peak at (87.4±0.76) days with a significance far above 5σ. Surprisingly, its position coincides with the phase shift analysis applied before. It is worth stressing that this independently derived result is also more rigorous, since it is based on the raw data. In addition, it demonstrates that the applied procedure with the daily interpolation is quite reliable. Then, the discussed possible involvement of some Mercury synod is actually redundant. Even the 0.5 day deviation (though within <1σ) from the 87.969 days orbital period of Mercury can be due to the scatter (σ=0.5 day) of the position in time of the monthly 12 mean values!

The other insert shows also a peak at (224.4±1.3) days, which coincides with the orbital period of Venus (224.47 days); this value has been derived also before. However, its significance (Figure 7c) is at the level of about 3.4σ.

*In short*, the Fourier analysis of the raw data confirms that the melanoma appearance shows also a short periodicity which strikingly coincides with the orbital period of Mercury.

## 3. Conclusions

The aim of this work was to establish planetary dependence in medical long time observations. In fact, the main result of this work is the observed fast oscillatory behavior of melanoma diagnoses (SEER monthly raw data 1973-2010), which should be a random distribution in the absence of any periodic impact. Astonishingly, a locked-in oscillation repeats coherently 11 times in 33 months, while it was summed-up 13 times over the observational time interval of 38 years. This is a highly significant signature (well above 5σ) of planetary relationship for the melanoma appearance. Phase shift analysis of the underlying basic oscillation (Figure 7a) using also its sign relative to a 3x3 months period following the measured monthly mean values, the orbital period of Mercury (or, in principle, some of its synodic period) is recovered. The phase shift reasoning of the raw data constrains the underlying period between about 87 and 90 days, coinciding with Mercury's orbital period or its synods. A more precise analysis provided a period of 87.5 days (subsection 2a.). In addition, in the longitudinal reference frame of Mercury (Figure 3), the observed amplitude of 2.75% supports the planetary observation, pointing at the orbital period of Mercury (=88 days) as a possible basic oscillation of the melanoma diagnoses. For the purpose of this work, it is of "minor" importance whether the oscillation of the melanoma rate is due to the Mercury's orbital period or some near synod.

Furthermore, we recall that Mercury has an about 100 times weaker dipole magnetic field than the Earth, while the Venus has actually no magnetic field. Gravitational or magnetic forces over planetary distances are extremely weak for explaining *somehow* any effect in living matter (see also ref.[1]). However, the observed Venus longitudinal spectral shape using linearly interpolated data (Figure 2) provides additionally an important piece of evidence in favour of the planetary relationship. Moreover, the derived rather narrow and/or peaking distributions in some of the combined planetary spectra (Figure 4) provides further supporting evidence for planetary relationship of melanoma appearance. As it has been argued for the physics case [1], peaking spectral shapes, exclude a long range force to be at the origin, which by its nature extends smoothly over an orbital period.

We note that the melanoma latency may be on the order of years or decades between initiation, onset, and detection. Though, the fact that the original monthly SEER data show an about 30% effect in 1 year over a time interval of 38 years, this indicates already that in about 30% of the melanoma diagnoses the latency cannot be (much) more than 1 year, otherwise the annual distribution gets smeared out accordingly. However, due to the observed periodicity near Mercury's orbital period (88 days), there must be also another latency of the order of 1 month for melanoma appearance. Exposure may have a cumulative effect where a final exposure crosses a threshold of detection. Because otherwise most derived plots in this work should be flat, i.e., random. In other words, this work answers in its own right the latency question.

*In summary*, this work, with Figures 2, 3 and 7, proves that a short latency (about one month) <u>must</u> exist in 1-10% of melanoma diagnoses, in order to explain the appearance of the derived periodicity of (87.5 ± 0.76) days. Following all the presented signatures combined along with the extremely weak gravitational / magnetic forces over the large planetary distances, the underlying scenario may be as in [1]: an occasionally gravitationally enhanced flux of streams of invisible massive matter interacts with the human body, surpassing a certain threshold. Remarkably, the observed peaking planetary orbital distributions including periods which do not coincide with some of the known orbital periods in the solar system (e.g. 156, 234, 440 days), or, the astrophysically inspired periods of unknown origin like the Rieger frequency (~156 days), make the present observations even more intriguing. Therefore, recovering new long time observations with more precise time resolution, e.g., 1 day, could be instrumental in establishing the nature of the assumed "invisible matter". Then, humans are the overlooked target and detector of the "Dark Universe" we are living in.

**Acknowledgments**

K.Z. wishes to thank Sergio Bertolucci and Horst Fischer for the years long collaboration in the field of astropartcicle physics. We profited a lot from the support by Antonis Gerdikiotis, Sebastian Hofmann and Marios Maroudas. Giovanni Cantatore and Yannis Semertzidis with their criticism, and final encouragement, helped arrive at this paper version. The support provided by CERN the last 2 decades is gratefully acknowledged making this spin-off possible. The fast response by the CERN librarians, and Tullio Basaglia in particular, was essential. We also thank Frédéric Clette from the Royal Observatory of Belgium, Brussels, who kindly provided the Sunspot data from the World Data Center SILSO. We also wish to thank Markus Aschwanden, who provided the large list of solar flares, which allowed us to derive Figure 5 in this work. Last but not least, the critical reports by both referees were suggestive in arriving to the present version of this work. The very first version from the 23$^{rd}$ March 2018 was not accepted by the arXives moderation. This reinforced us to go deeper and arrive finally to this version, and we are thankful for this!

**FIGURES**

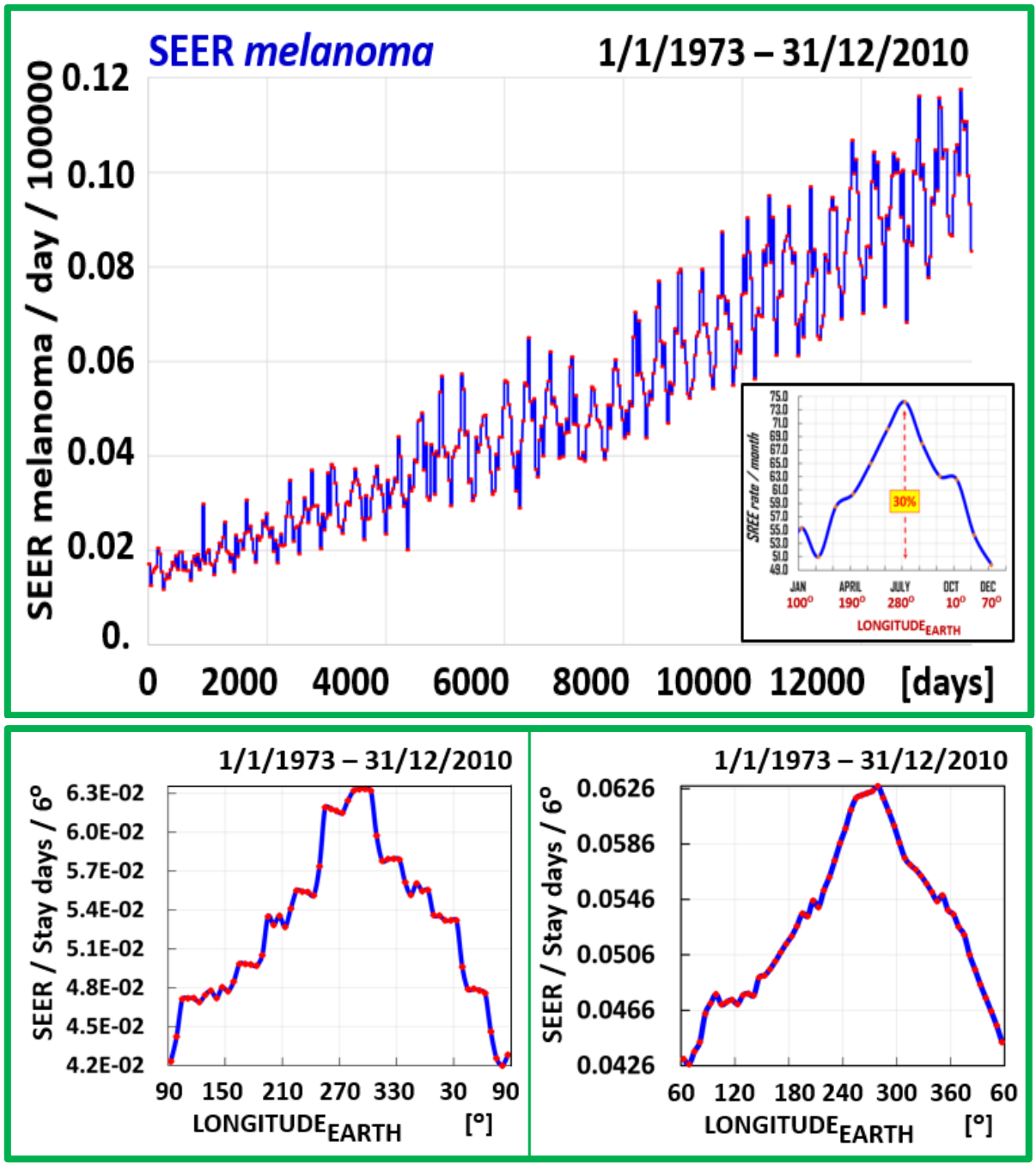

**Figure 1.** UP: The estimated daily rate of diagnosed **melanoma** for a period of 38 years (1/1/1973 - 31/12/2010) normalised to a population of 100000 people [2]. In the insert is shown the sum of the monthly mean values during the 38 years long observation. The peak in July corresponds to about 280º in Earth longitude. The striking oscillations seen (blue line) are mainly the annual modulations. DOWN: The daily melanoma rate is estimated from each monthly mean value divided by the length of the month (*left*). A linear interpolation between neighboring monthly values gives a more smooth distribution (*right*). The so derived day of the melanoma rate projected on the Earth's longitude gives the longitudinal distributions.

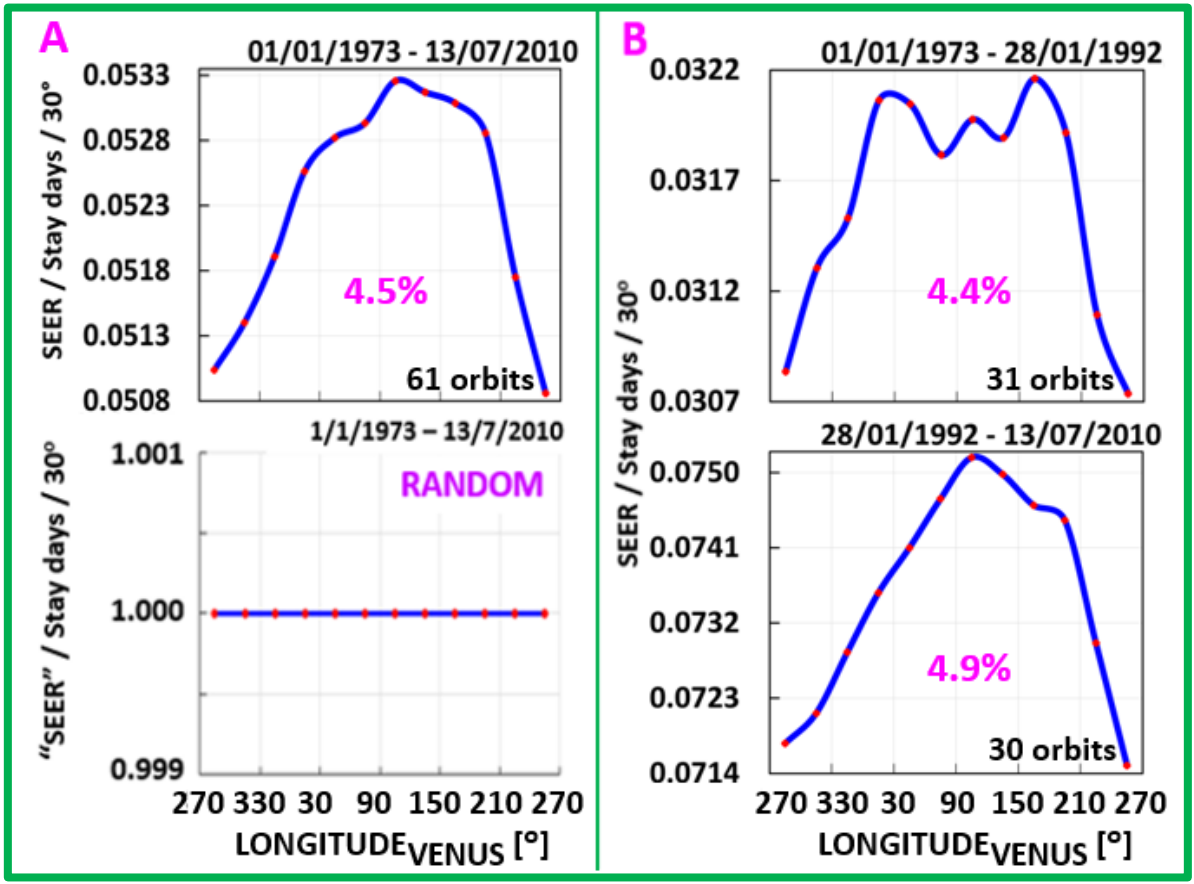

**Figure 2.** The projected melanoma rate on the longitudinal distribution of **Venus** using the daily melanoma rate [2] for BIN=30° (linear interpolation). The spectra of the whole observation period (A) and the corresponding two halves (B) are shown. For these plots no other planetary constraint was applied. To crosscheck the applied evaluation procedure, the same evaluation code was running by using a constant random event rate of 1/day; the result is, as expected, a completely flat distribution at 1/day (the lower spectrum in column A). The difference between maximum and minimum value is given in percent for each spectrum. (About the possible phase shift between the two halves in (B) see subsection 2a.).

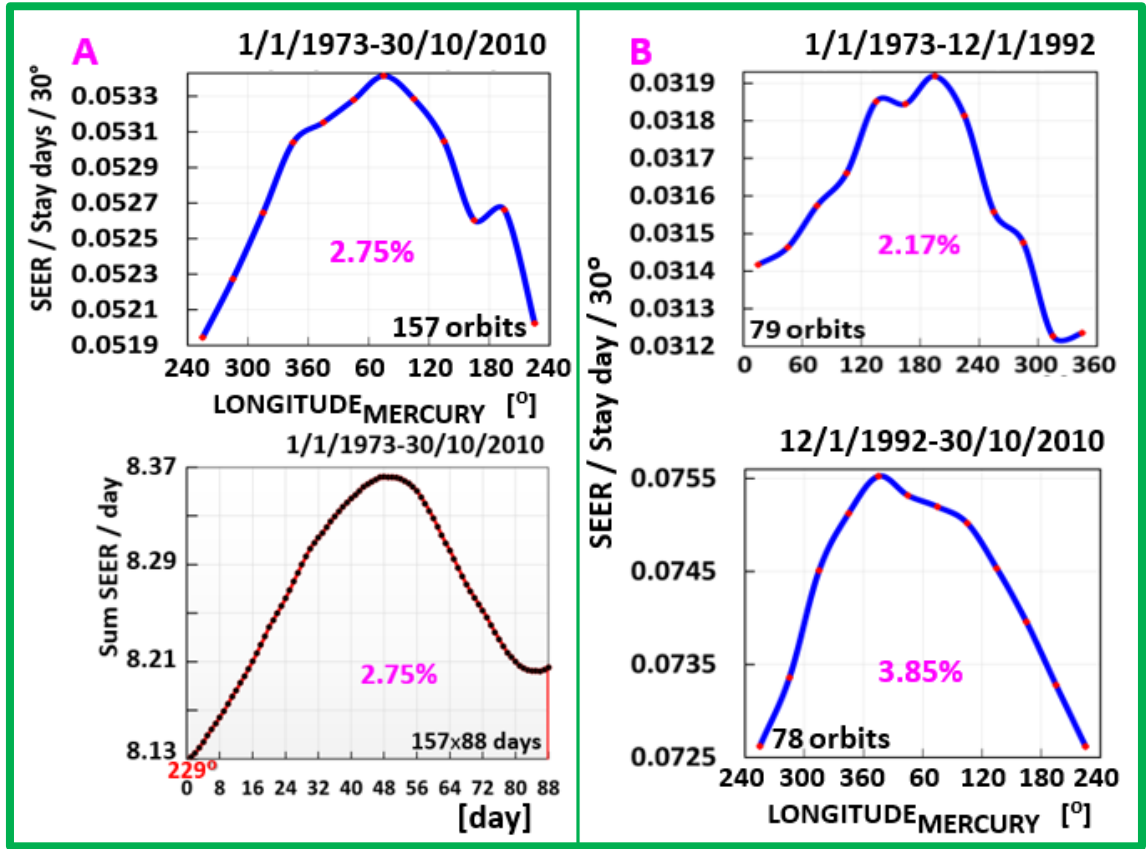

**Figure 3.** The projected melanoma rate on the longitudinal distribution of **Mercury** using the daily melanoma rate [2] for BIN=30º (linear interpolation). The spectra of the whole observation period (A) and the corresponding two halves (B) are shown. For comparison, it is shown (A lower spectrum) the sum of the estimated daily melanoma rate of 157 consecutive time intervals of **88 days**. For all plots no other planetary constraint was applied. For the observed phase shift between the two halves in (B) see subsection 2a.. The difference between maximum and minimum value is given in percent for each spectrum.

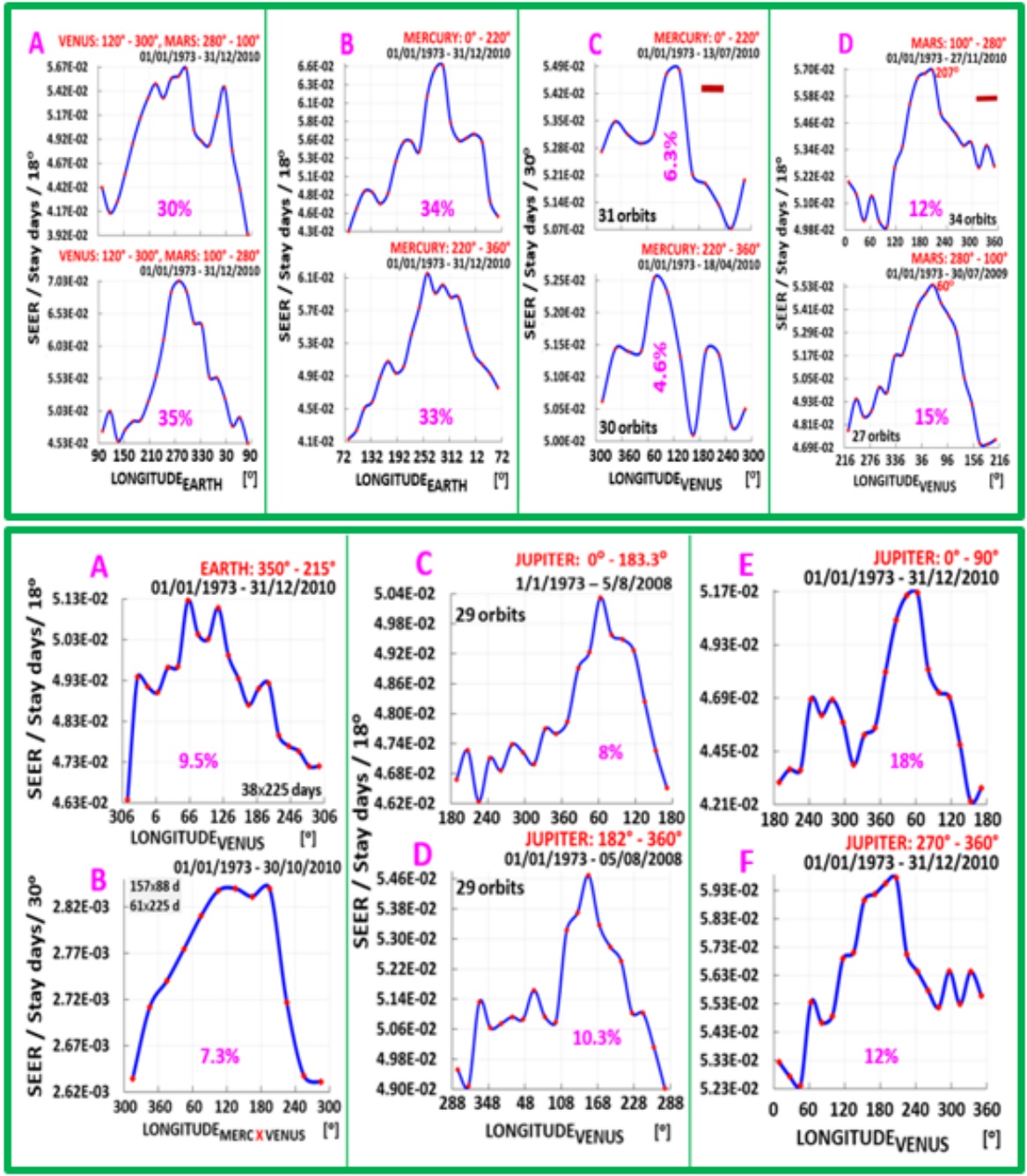

**Figure 4.** UP: The projected melanoma rate [2] longitudinal distributions for the planets **Earth** (A,B), and **Venus** (C,D) applying additional planetary constraints given in red. To compare a pair, notice possible differences in spectral shape and amplitude as well as the longitudinal position of a peaking distribution. DOWN: In B is shown the melanoma rate longitudinal distribution by taking the product BIN by BIN of the spectra of Mercury and Venus. The difference between maximum and minimum value is given for each spectrum in percent.

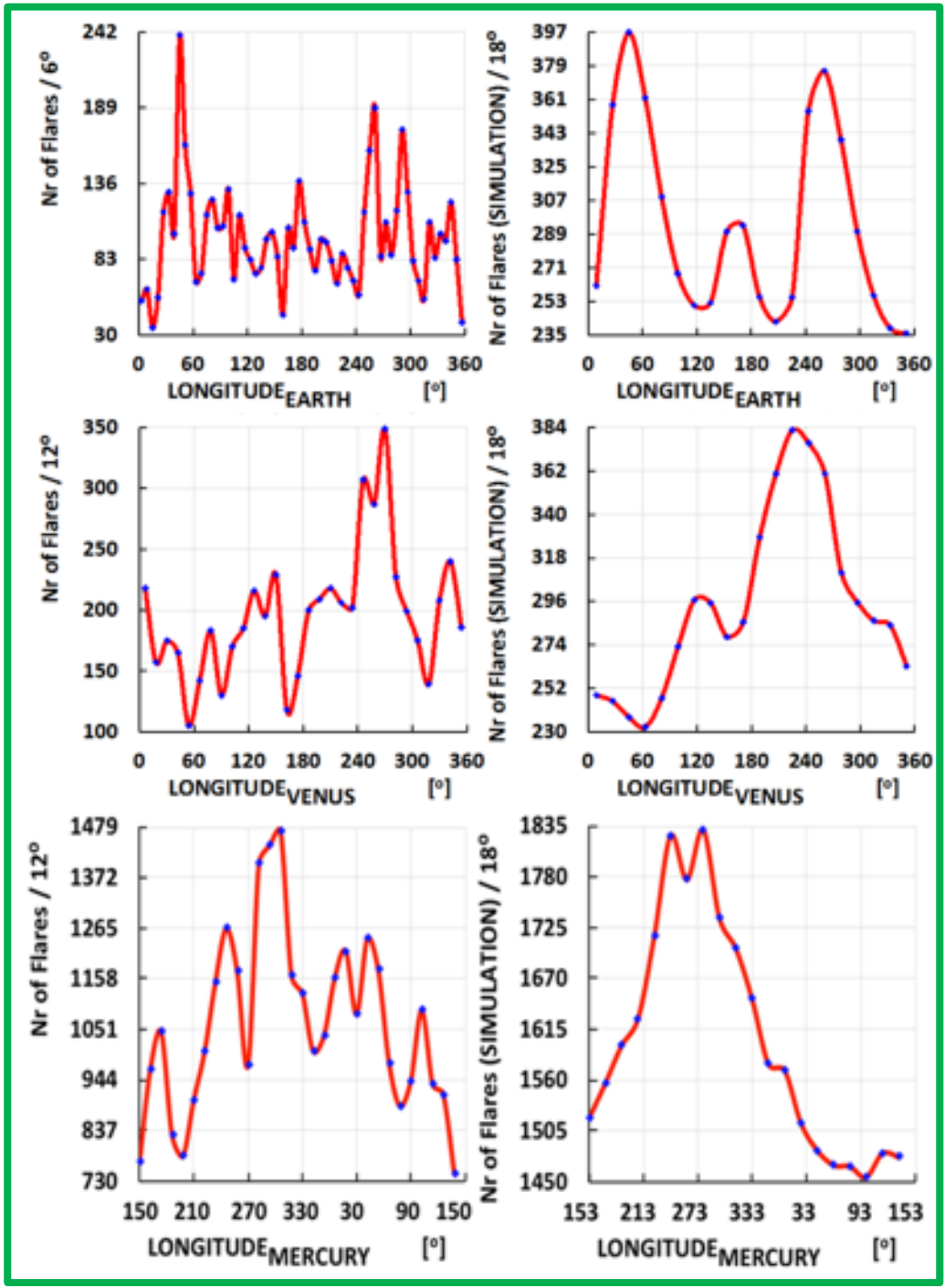

**Figure 5.** *Left*: Longitudinal distribution of daily solar flares (1972-2015) [5]. Their day of appearance is projected on the orbital position of the inner 3 planets. One or more peaks are seen in each spectrum (see ref. [1]). *Right*: The number of flares of 30 consecutive days are summed-up, simulating the monthly rate. Then, with a linear interpolation between neighboring monthly values (while dividing by the number of days of each month), the simulated daily distribution of flares is estimated. The whole procedure resembles the treatment of the SEER data. Running the same analysis code as for the real data (*left*) the simulated spectral distributions are derived (*right*). The degree of distortion of the original spectral shapes is acceptable for the purpose of this work; Mercury's decreasing signal-to-noise ratio is the largest (~2.5×). This simulation performed Marios Maroudas / University of Patras.

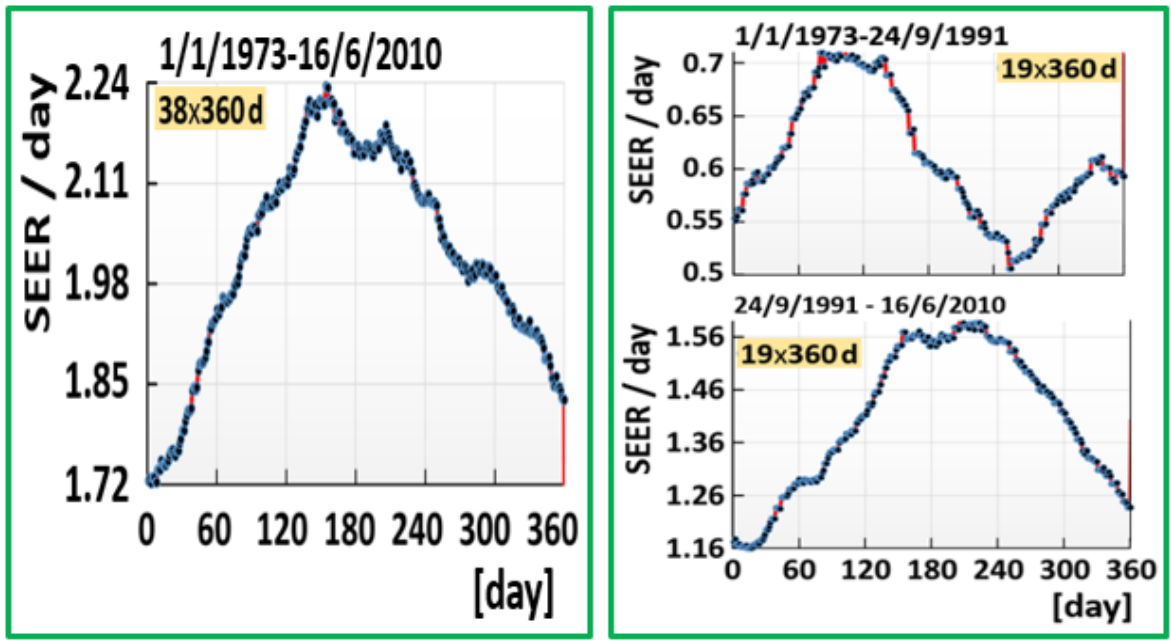

**Figure 6.** *LEFT*: The sum of the estimated daily melanoma rate [2] of 38 consecutive time intervals of **360 days** [2], which is near the Earth's annual periodicity (365.25 days). *RIGHT*: The melanoma spectrum on the left has been estimated for two partial phase-locked equal long time intervals. In order to follow-up the evolution with time of the spectral shape, the same computer output was used to derive all three spectra. The observed phase-shift over the 19 years between both spectra is about 105 days. This is consistent with the accumulated phase difference between the two periods 19 years apart: 19×(365.25-360) ≈100 days. (See also Section 2b).

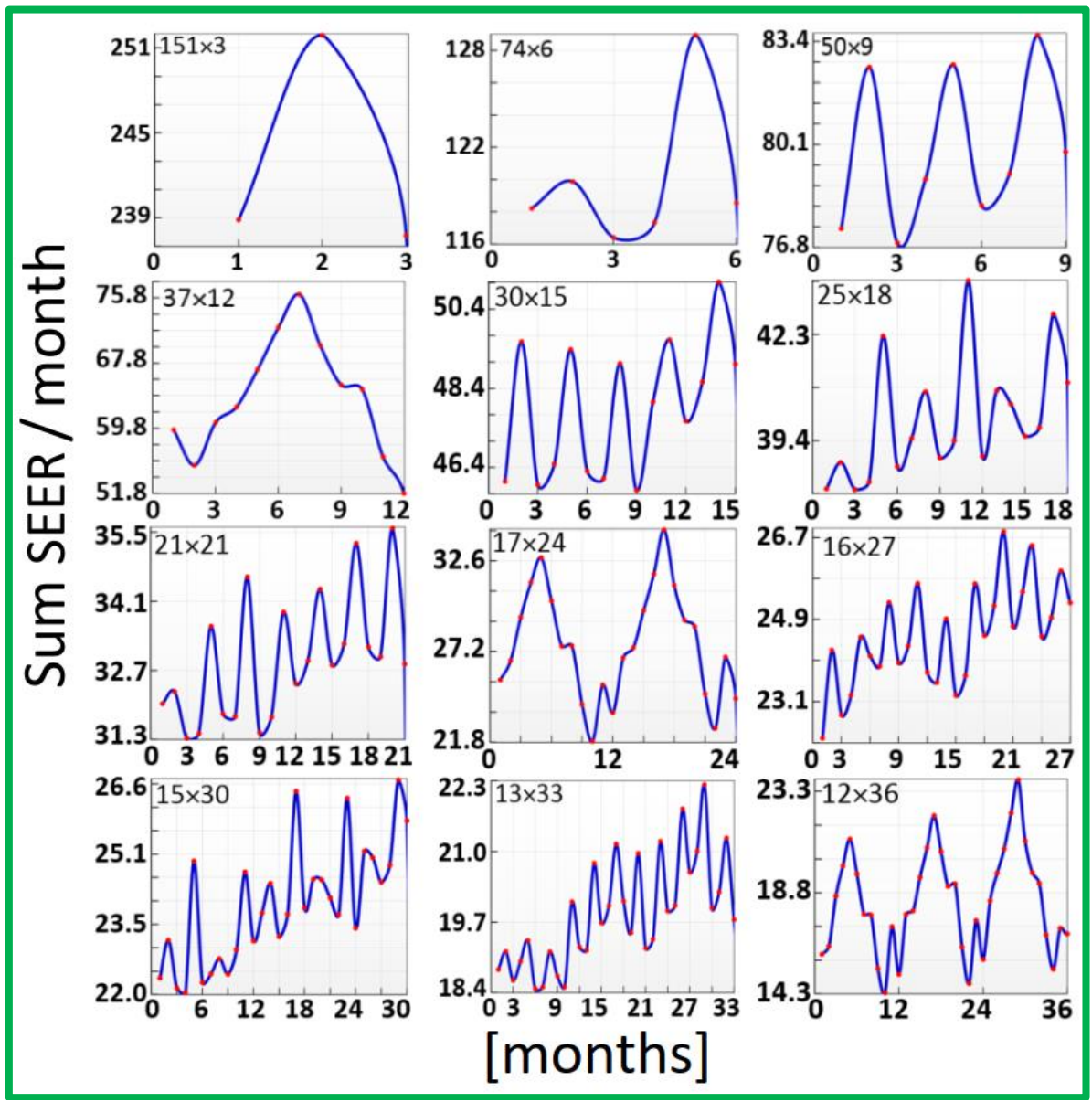

**Figure 7a.** The sum of the original monthly melanoma diagnoses for the period starting 1/3/1973 up to 31/12/2010 [2] of 12 different consecutive time intervals **3 - 36 months**. A 'fundamental' 3 months period does oscillate coherently up to 11 times in 33 months. Note, time intervals of 12, 24 and 36 months show the annual periodicity and its multiples.

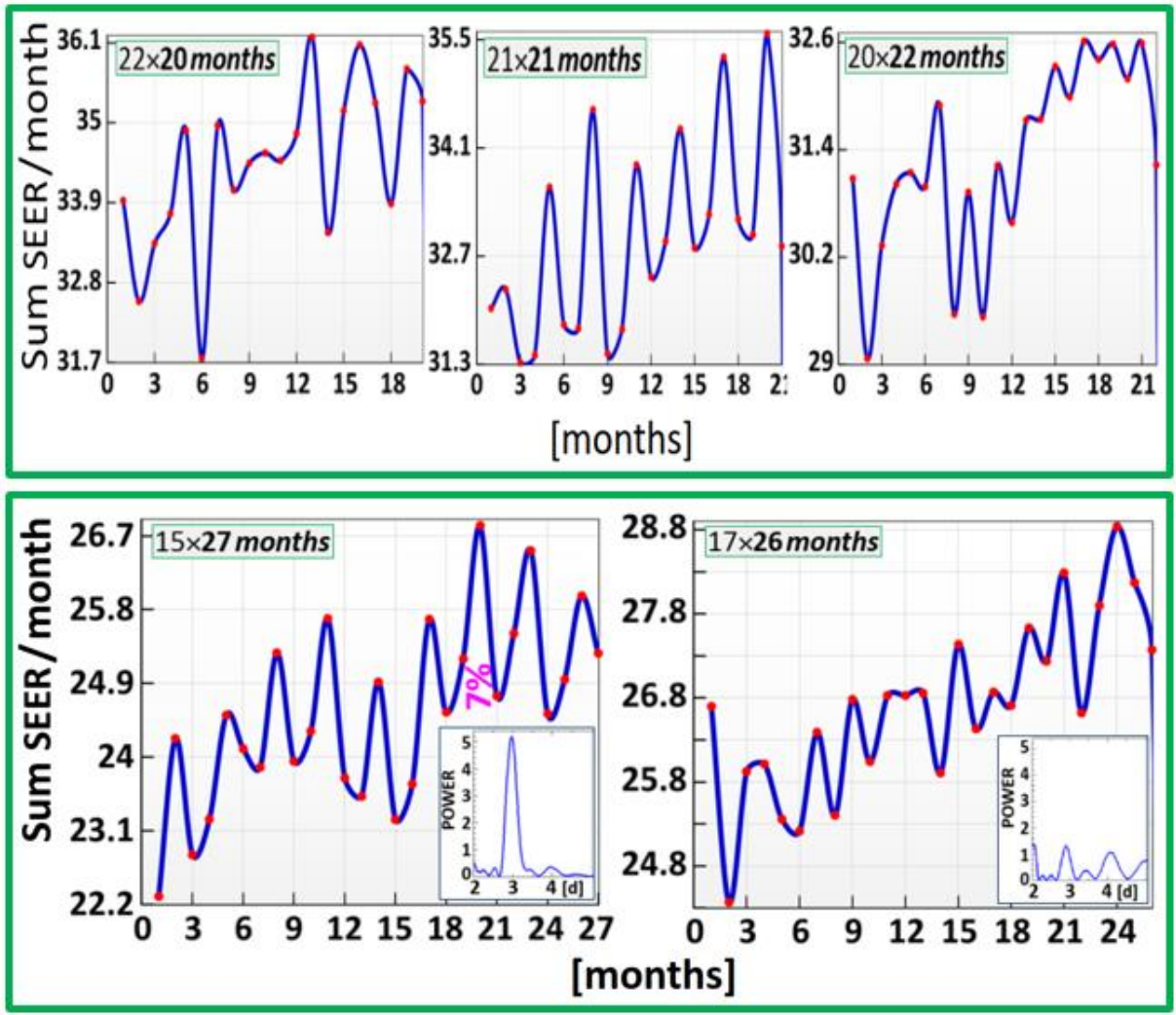

**Figure 7b.** Two periodic spectra are taken from Figure 7a for **21 & 27 months**. The upper frame provides a comparison with the two neighboring spectra of +/- 1 month time interval. The 7 times repeating "resonance" of 3 months disappears in the neighboring ones. In the lower frame it is shown the 9 times repeating "resonance" at the 3 months 'fundamental' mode; the insert shows a dominating Fourier peak of 3 months. For comparison it is shown also the near time interval of 26 months, whose apparent irregular oscillatory behavior is seen also in the Fourier insert. Assuming Poisson statistics, the relative error per point is <0.7%. The estimated significance of the 7% amplitude is **7σ**.

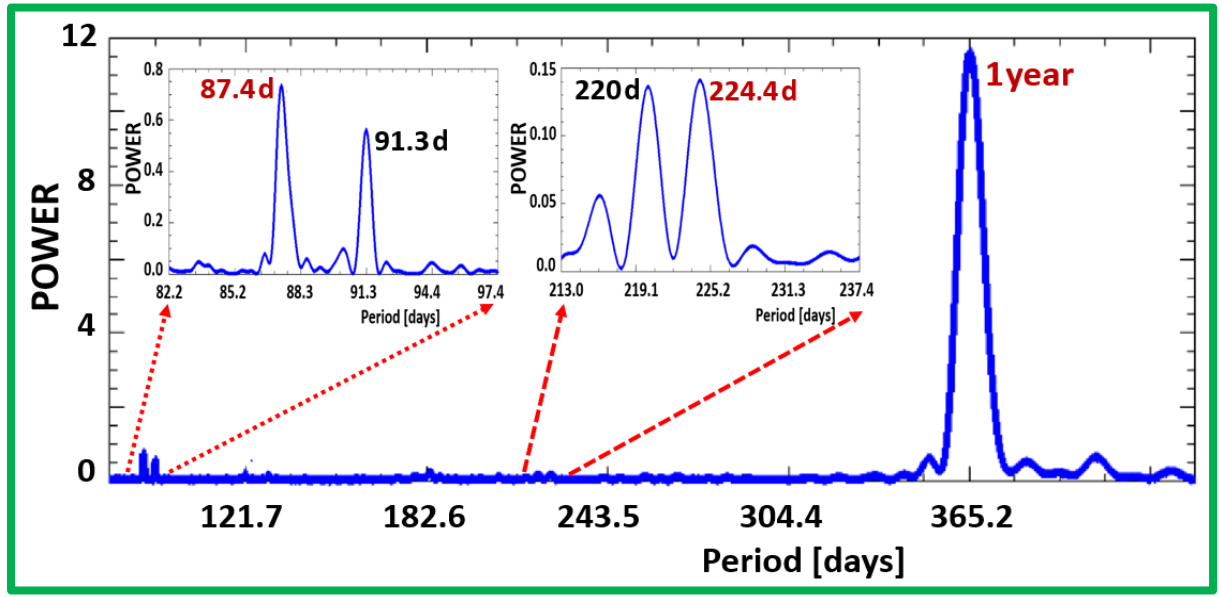

**Figure 7c.** The Fourier analysis of the original 456 monthly SEER data (1973-2010). The dominating annual period is seen also in the "light curve" (Figure 1). Near the 4$^{th}$ harmonic of the main annual peak at 91.3 days (*left insert*) appears an even stronger peak at (87.4 ± .76) days with a significance far above **5σ**. Remarkably, its value coincides with the orbital period of Mercury (=87.969 days). The peak at (224.4 ± 1.3) days (*right insert*) coincides with the orbital period of Venus (224.47 days), but its significance is about 3.4σ.
Fourier analysis was prepared by Antonis Gardikiotis / University of Patras.

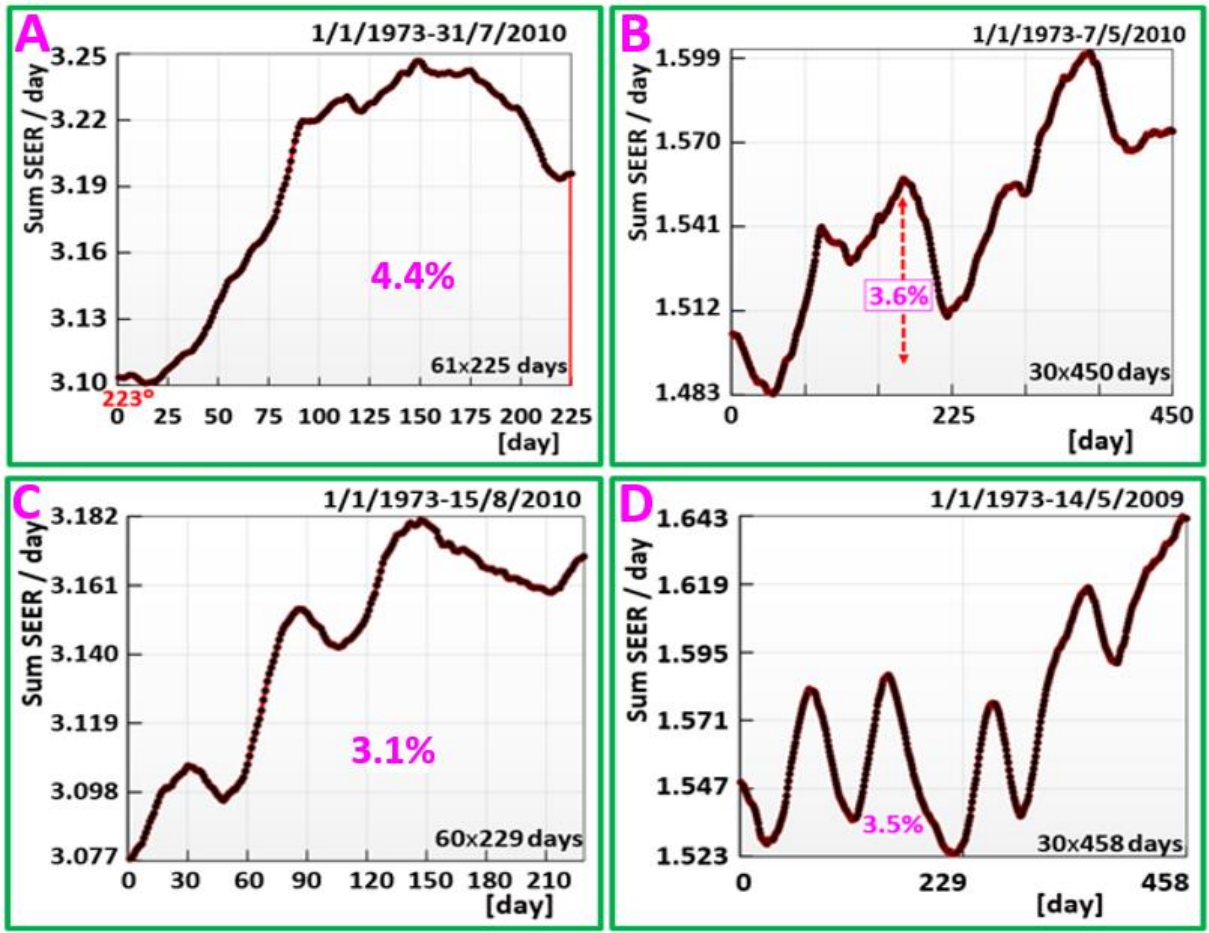

**Figure 8.** The sum of the daily melanoma rate (following from linear interpolation) [2] of 61 consecutive time intervals of **225 days** (A) is near the orbital period of Venus (= 224.47 days). Similarly, the distribution in (B) arrives from integrating 30 consecutive time intervals of 2×225=**450 days**. The same analysis gives the spectra C and D by using as time interval **229 days** and 2×229=**458 days**, respectively. Note, 229 days is the synodic period of Venus-Saturn. Raw melanoma data with much shorter time BIN than 1 month, can definitely clarify this possible indication in C and D. The nearby periods of 225, 229 and 237 days (Venus' largest synodic period) could result to some unforeseen interference effects.

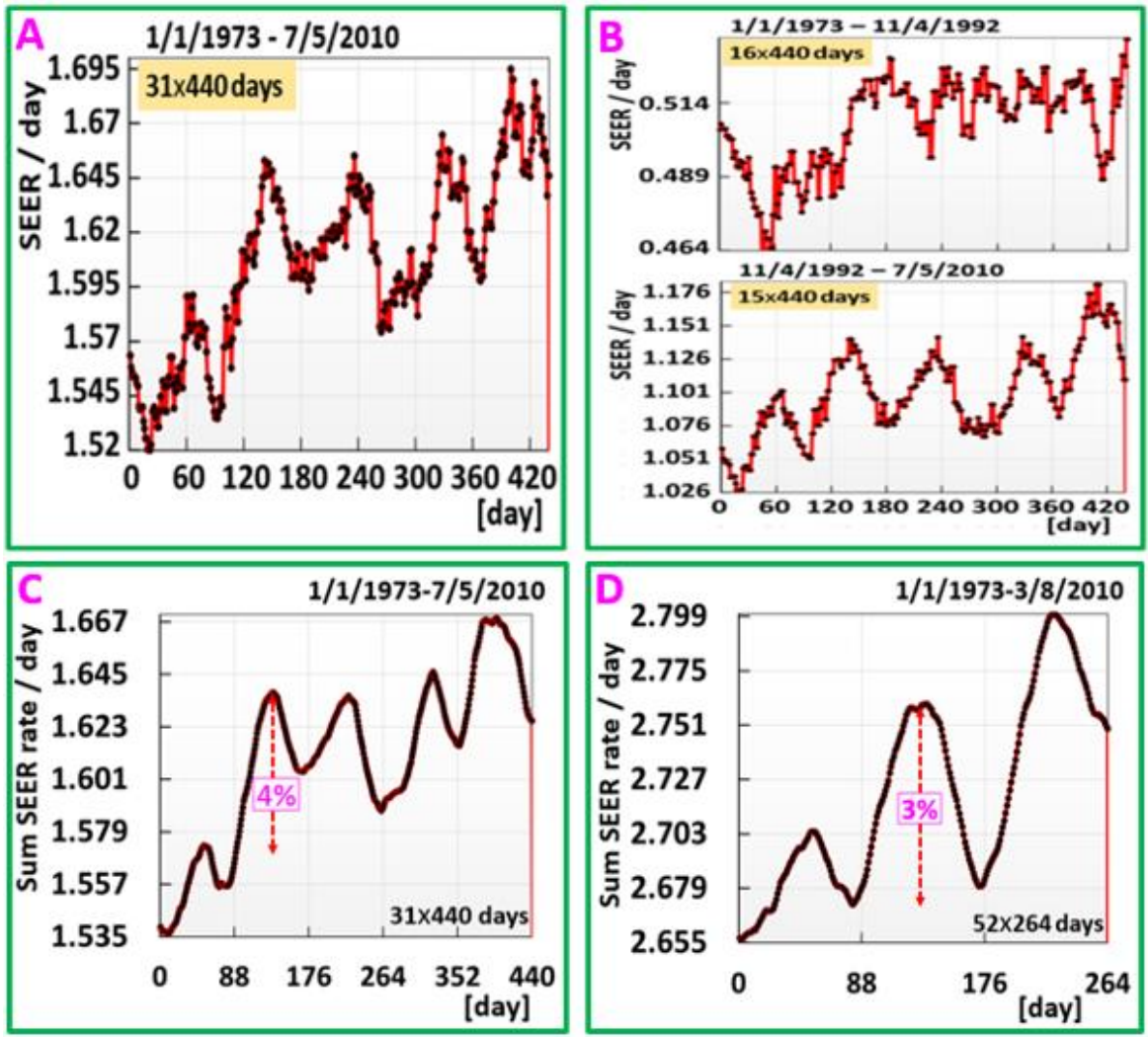

**Figure 9.** The sum of the reconstructed daily melanoma rate [2] of 31 consecutive time intervals (A) of **440 days**, which coincides within 4 hours with the sum of 5 orbital periods of Mercury. The spectrum (C) is similar to (A), but the daily rate has been calculated from the monthly values by linear interpolation. The 5 peaks in A, B and C reflect the 88 days periodicity of Mercury (= 87.969 days). Fourier analysis gives 88.3 days for (A) and 87.9 days for (C) [7]. The double plots in (B) are from two equal long time intervals. The second half (B, LOW) shows a striking 88 days modulation, which in the first half (B, UP) it appears as one 91.6 days component in the Fourier power spectrum [7]. The spectrum (D) is derived as (C), but adding-up 52 consecutive time intervals of 3×88=**264 days**, i.e., 3 times the orbital period of Mercury. Spectra A & C and D should be compared with the corresponding ones in Figure 7a 30×15 & 50×9, respectively. This comparison is a quality test for both approximations used in this work to derive a daily melanoma rate.

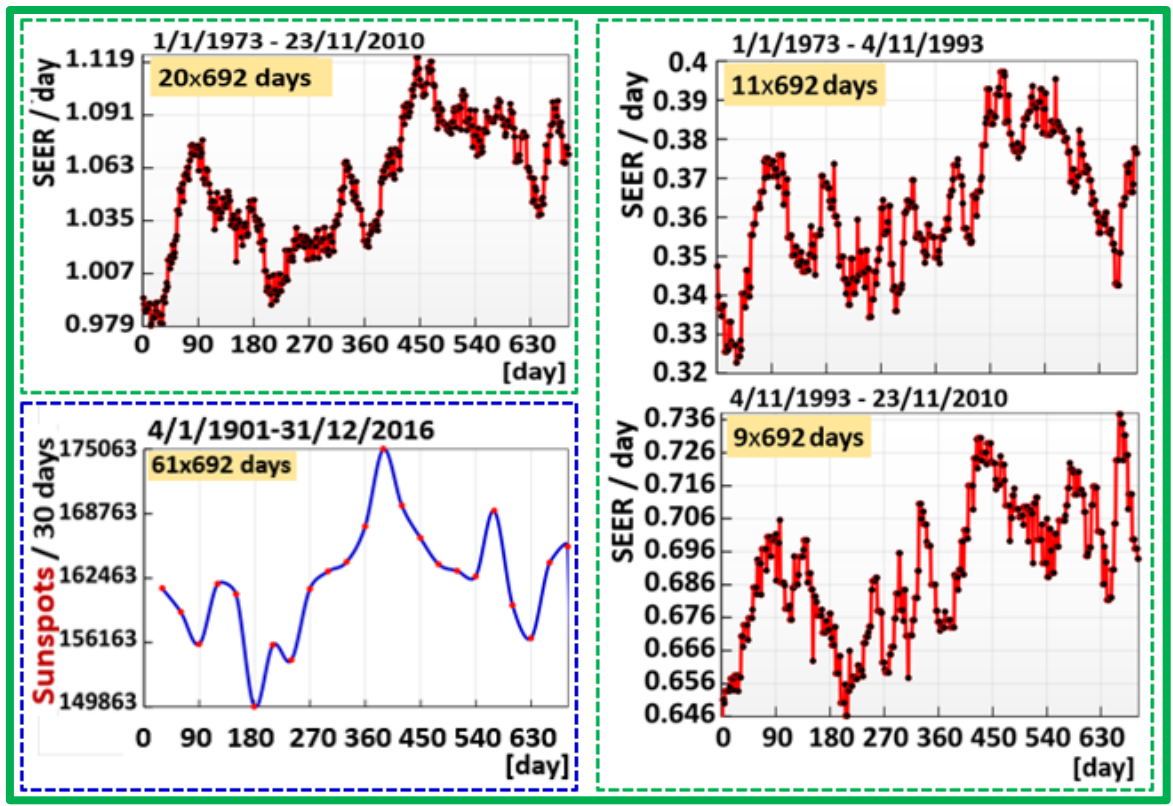

**Figure 10.** *LEFT*: The sum of the daily melanoma rate [2] of 20 consecutive time intervals of 692 days, which is equal to the **Mars-Pluto synod**. (DOWN) A similarly derived spectrum with sunspots [6] shows a large peak amplitude (>14%), which was the reason in this work for choosing this time interval. *RIGHT*: The two spectra are from two almost equal long time intervals, splitting the melanoma spectrum on the left in two halves. In order to follow-up the spectral shape in time the same computer output was used in deriving all three melanoma spectra, indicating a negligible phase shift between both halves. Using the narrow peak at the end of both spectra the possible phase-shift is negligibly small (~ 10 days).

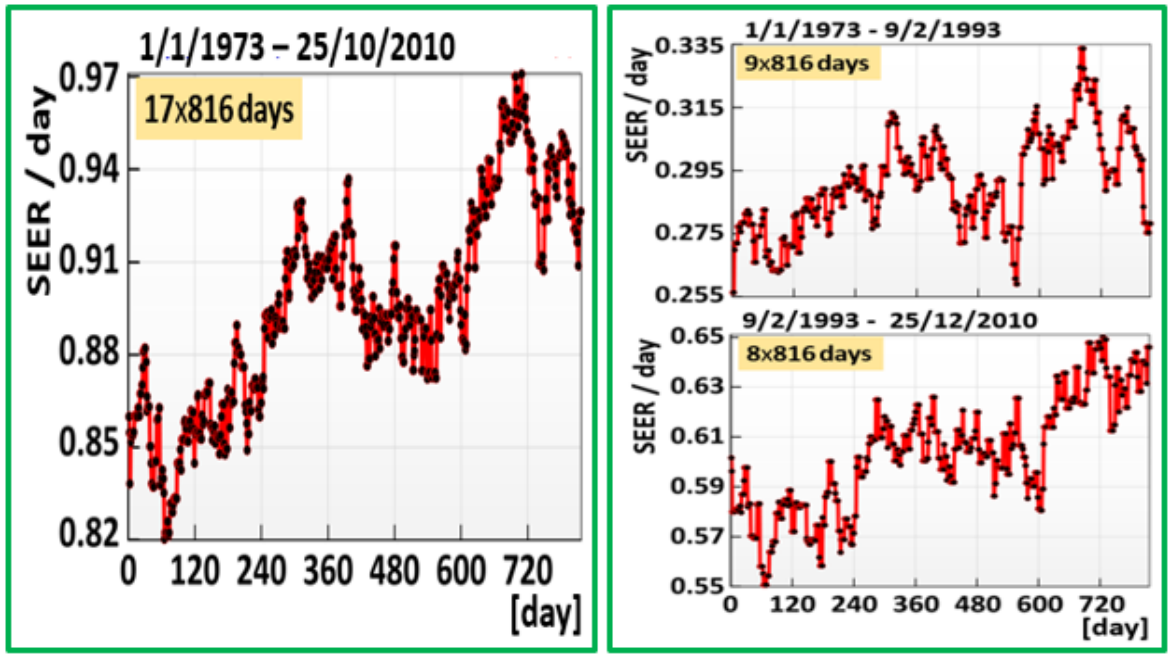

**Figure 11.** *LEFT*: The sum of the daily melanoma rate [2] of 17 consecutive time intervals of 816 days, which is equal to the **Mars-Jupiter synod**. *RIGHT*: The two spectra are from two almost equal long time intervals, splitting the melanoma spectrum on the left in two halves. In order to follow-up the spectral shape in time the same computer output was used in deriving all three spectra. A relative phase shift between both halves cannot be deduced. This may be the reason why the two broad peaks appear more pronounced in the total spectrum on the left.

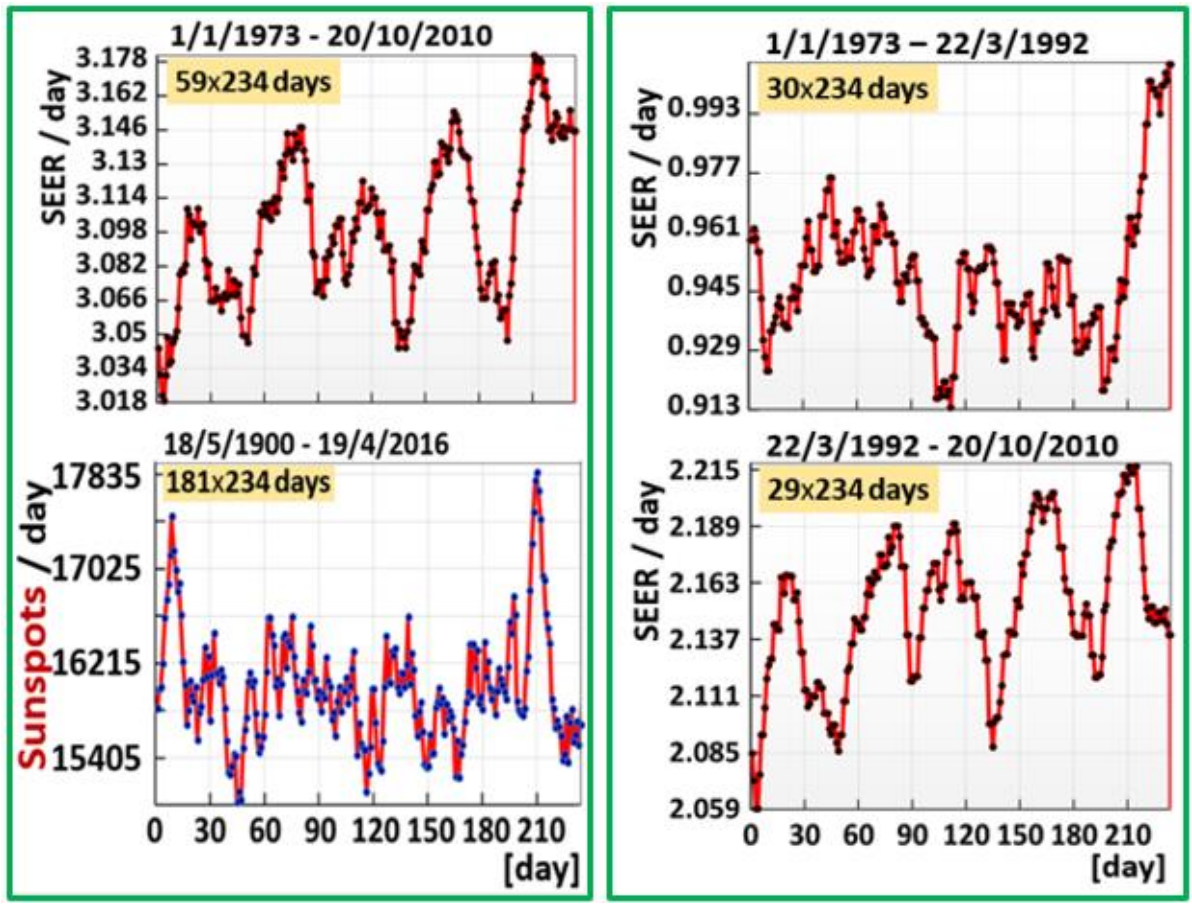

**Figure 12.** *LEFT*: (UP) The sum of the daily melanoma rate [2] of 59 consecutive time intervals of **234 days** shows a short oscillatory behavior of about 47 days. (DOWN) A similarly derived spectrum with the sunspots [6] shows remarkably two peaks, and this was the reason to analyze this particular time interval in this work. *RIGHT*: The two spectra are from two almost equal long time intervals, splitting the melanoma spectrum on the left in two halves. In order to follow-up the spectral shape in time the same computer output was used in deriving all three spectra. The second half shows a striking ~47 days modulation. A similar behavior is obtained using linearly interpolated daily SEER data (not shown). The general behavior seems similar to that of Figure 9 with the 88 days oscillation. Though, only a similar spectrum, with raw melanoma data and much shorter time BIN than 1 month, can definitely clarify this possible indication. Because, the nearby periods between 225 days (Venus' orbit) and 237 days (largest Venus synod) could result to some unforeseen interference effects.

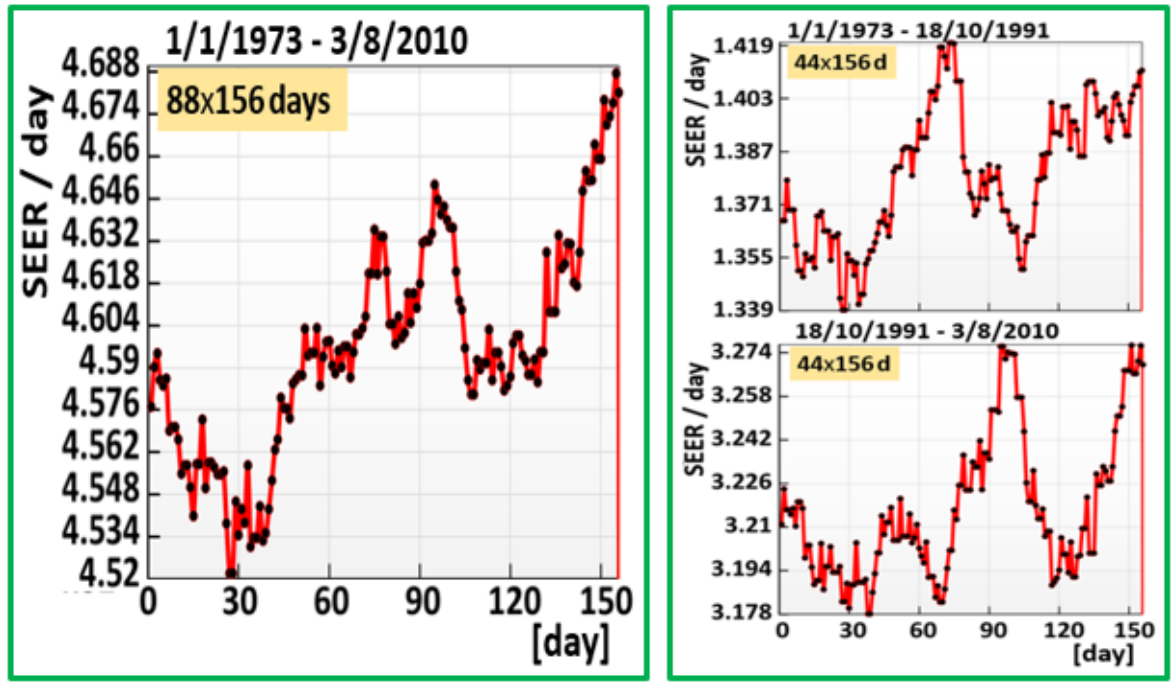

**Figure 13.** *LEFT*: The sum of the daily melanoma rate [2] of 88 consecutive time intervals of 156 days. This time interval is equal to the "**Rieger period**" observed with energetic solar flares [8], whose origin is unknown. *RIGHT*: The two spectra are from two equal long time intervals, splitting the melanoma spectrum on the left in two halves. In order to follow-up the spectral shape in time the same computer output was used in deriving all three spectra. In each half the observed striking peaks are phase shifted relative to each other by about 50 days. This is understandable taking into account that the Rieger periodicity is not precisely known, while interference effects with some nearby period can also cause a phase shift. As crosschecking, a similar behavior is obtained with linearly interpolated daily SEER data.